\documentstyle[prl,aps,epsf,multicol]{revtex}
\begin{document}
\title{\bf Survival Probability of  a Ballistic Tracer Particle in the
Presence of Diffusing Traps}
\author{Satya N. Majumdar$^1$ and Alan J. Bray$^{1,2}$}
\address{$^1$Laboratoire de Physique Quantique (UMR C5626 du CNRS), 
Universit'e Paul Sabatier, 31062 Toulouse Cedex, France \\
$^2$Department of Physics and Astronomy, University of 
Manchester, Manchester M13 9PL, UK}

\date{\today}

\maketitle

\begin{abstract}
We calculate  the survival  probability $P_S(t)$ up  to time $t$  of a
tracer  particle   moving  along  a  deterministic   trajectory  in  a
continuous  $d$-dimensional space  in  the presence  of diffusing  but
mutually noninteracting  traps.  In particular, for  a tracer particle
moving ballistically with a constant  velocity $c$, we obtain an exact
expression for $P_S(t)$, valid for  all $t$, for $d<2$. For $d\geq 2$,
we obtain the  leading asymptotic behavior of $P_S(t)$  for large $t$.
In all cases, $P_S(t)$ decays exponentially for large $t$, $P_S(t)\sim
\exp(-\theta  t)$. We  provide  an explicit  exact  expression of  the
exponent  $\theta$ in  dimensions  $d\leq 2$  and  for the  physically
relevant case, $d=3$, as a function of the system parameters.

\noindent

\end{abstract}

\begin{multicols}{2}

The  calculation of  the  survival probability  of  a tracer  particle
moving  in  the presence  of  diffusing traps  is  a  problem of  long
standing interest as it appears,  in various guises, in a wide variety
of  contexts such  as reaction-diffusion  systems\cite{Rice}, chemical
kinetics\cite{Benson,OZ,TW}, predator-prey models\cite{PP} and `walker
persistence'  problems\cite{WP}.  The  tracer particle  dies instantly
upon meeting any of the  diffusing traps.  Perhaps the simplest of all
these  problems  is  the  case  when  the  diffusing  traps  are  {\em
noninteracting} and the  motion of the tracer particle  is governed by
its own intrinsic  dynamics that depends on the  specific problem. For
example, if  the tracer particle is  static, this problem  is known as
the {\em target annihilation} problem\cite{TA}. Of particular interest
is the case when the tracer  particle itself has a diffusive motion, a
problem that  was first studied  by Bramson and  Lebowitz\cite{BL} and
has recently seen a flurry of activity\cite{MG,BB1,BB2,OBCM,BMB}. It is,
however, somewhat frustrating  that, despite various new developments,
this  {\em diffusive target  annihilation} problem  has defied  a {\em
direct} exact solution.   In contrast, we show in  this paper that the
{\em ballistic target annihilation} problem, where the tracer particle
moves ballistically with a constant velocity, is exactly solvable.

A precise definition of the general problem is as follows.  Consider a
set of particles  initially (at time $t=0$) distributed  randomly in a
continuous $d$-dimensional space with average density $\rho$.  Each of
these  particles subsequently  undergoes independent  diffusive motion
with the same diffusion constant  $D$. A tracer particle is introduced
into  the  system  at  $t=0$  at the  origin  and  subsequently  moves
according to its own prescribed equation of motion. This motion can be
either deterministic  or stochastic, depending  on the problem.  For a
given trajectory ${\vec R_0} (t)$ of the tracer particle, we ask: what
is the probability  $P_S(t)$ that none of the  random walkers hits the
tracer particle up to time $t$?  Evidently $P_S(t)$ depends implicitly
on the trajectory  ${\vec R_0}(t)$. For a deterministic  motion of the
tracer particle  where the  trajectory ${\vec R_0}(t)$  is prescribed,
its survival probability is precisely $P_S(t)$. On the other hand, for
stochastic motion  of the tracer particle the  survival probability is
obtained  by   subsequently  averaging  $P_S(t)$   over  all  possible
trajectories of the  tracer particle.  Thus the basic  step, in either
case,  is to  compute $P_S(t)$  for a  given  deterministic trajectory
${\vec R_0}(t)$.   In this  paper we limit  ourselves to the  study of
this  general  deterministic trajectory  problem  and, in  particular,
present explicit  exact results when the trajectory  is ballistic (see
Fig. 1).
\begin{figure}
  \narrowtext\centerline{\epsfxsize\columnwidth \epsfbox{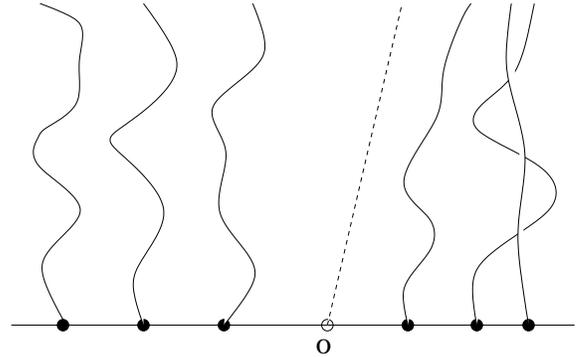}}
\caption{A   schematic  picture  of   the  space   (horizontal)-  time
(vertical) trajectories of the diffusing traps (solid curves) and that
of the ballistic tracer particle (dashed line) in $d=1$. The diffusing
trajectories are allowed to cross each other.}
\end{figure}

For  a given  fixed trajectory  ${\vec R_0}(t)$,  this problem  can be
formally reduced to a single-particle problem as follows. Let $Q(\vec
r_i, t)$  be the probability  that the $i$-th random  walker, starting
from the  initial position  $\vec r_i$, does  not meet  the trajectory
${\vec R_0}(t)$ up to time $t$.  Note  that  $Q(\vec  r_i,t)$  depends
implicitly  on the  trajectory ${\vec  R_0}(t)$. The  initial position
$\vec r_i$ of the $i$-th  walker is a random variable distributed over
a  volume $V$  of a  $d$-dimensional  space  with uniform  probability
density $1/V$. We assume that  there are $N$ such walkers.  Eventually
we  will  take  the  limit $N\to \infty$, $V\to  \infty$  keeping  the
density $\rho=N/V$  fixed. Since  the random walkers  are independent,
one    immediately   gets    $P_S(t)=\langle    \prod_{i=1}^N   Q(\vec
r_i,t)\rangle $ where  the angle brackets denote the  average over the
initial positions  $\vec r_i$   of  the random walkers. We  next write
$Q(\vec r_i,  t)=1-P(\vec r_i,t)$ and  use the product measure  of the
initial  condition to  write, $P_S(t)  = \left[  1 -  {1\over {V}}\int
P(\vec  r,  t) d\vec  r\right]^N$.   Taking  the  $V\to \infty$  limit
at fixed $\rho$ gives 
\begin{equation}
P_S(t)    =   \exp\left[-    \rho   \int    P(\vec   r,    t)d\vec   r
\right]=\exp[-\mu(t)],
\label{pst2}
\end{equation}
where $\mu(t)= \rho\int P(\vec r,t)\, d{\vec r}$ and $P(\vec r, t)$ is
the probability that a single  random walker starting at $\vec r$ hits
the  trajectory  ${\vec  R_0}(t)$  before  time  $t$.  Note  that  the
calculation  of  $P(\vec r,t)$  and  hence  that  of $\mu(t)$  is,  in
general, very  hard even for this  single particle problem  due to the
{\em  moving}  boundary  ${\vec  R_0}(t)$.  Such  problems,  generally
termed  Stefan   problems,  are  well   known  to  be   formidable  to
solve\cite{Sal}.  In some  simple  cases one can derive the asymptotic 
behaviors of $P(\vec  r,t)$ for large $t$\cite{PP}. Unfortunately, this 
asymptotic form of $P(\vec r,t)$ can not be used to perform the integral 
over $\vec r$ in Eq.\ (\ref{int1}) since this integral usually  diverges. 
One therefore needs an alternative approach.

Fortunately,  a  technique  for   computing  $\mu(t)$  for  a  general
trajectory ${\vec R_0} (t)$ has recently been developed \cite{BMB}. It
was  shown that,  for $d<  2$, $\mu(t)$  satisfies the  exact integral
equation\cite{BMB}
\begin{equation}
\rho  = \int_0^{t}  dt' {\dot  \mu}(t') G\left({\vec  R_0}(t), t|{\vec
R_0}(t'),t'\right),
\label{int1}
\end{equation} 
where   ${\dot  \mu}=d\mu/dt$   and  $G\left({\vec   R_0}(t),  t|{\vec
R_0}(t'),t'\right)=[4\pi  D(t-t')]^{-d/2}  \exp\{-[{\vec R_0}(t)-{\vec
R_0}(t')]^2/{4D(t-t')}\}$  is the  standard $d$-dimensional diffusion
propagator.  For the  marginal dimension  $d=2$, Eq.\  (\ref{int1}) is
still valid, though one has to introduce an ultraviolet cut-off in the
diffusion propagator.   The derivation  of this integral  equation has
been detailed in Ref. \cite{BMB}.  Note that for continuous space with
$d>2$, if the tracer is a point particle  then the random walkers will
never meet the  point particle trajectory. This is,  however, not true
on a lattice.  So, if one sticks to a continuous space, the problem is
sensible for $d>2$ provided the tracer particle has a finite size. The
formula in  Eq.\ (\ref{int1}) assumes a point  particle trajectory and
hence is valid  only for $d\leq 2$. For $d>2$,  there is no equivalent
formula and one has to use other methods (see later in the paper).

For a general trajectory ${\vec R_0}(t)$  it is not easy to invert the
integral  equation (\ref{int1})  to obtain an explicit  expression for
$\mu(t)$.  However, the advantage of Eq.\ (\ref{int1}) is that in many
cases  it can  be exploited  to derive  exact asymptotic  results. For
example,  for  the   `diffusive  target  annihilation'  problem,  Eq.\
(\ref{int1}) has been used  to derive asymptotically exact bounds
for  the survival  probability\cite{BMB}. Note,  however, that  in the
diffusive case, one had to average over all trajectories of the tracer
particle weighted with the Wiener measure\cite{BMB}.  Another solvable
case is in  $d=1$ when the tracer particle  moves deterministically as
$R_0(t)=c{\sqrt t}$. For this case an exact expression for $\mu(t)$ for 
large $t$ was obtained\cite{BMB}.

The purpose of this paper is to present another solvable case where the
tracer particle moves ballistically through the system. For $d\leq 2$,
we will consider  the tracer to be a point  particle with a trajectory
${\vec R_0}(t)= ct  {\bf {\hat z}}$, where $c$  denotes the velocity of
the particle  and ${\bf {\hat z}}$  denotes the unit  vector along the
direction of  motion, which we choose to be the  $z$ axis. For  $d>2$, 
we will consider the tracer particle to be a ball of finite radius $a$.
For $d\leq 2$,  we show how  Eq.\ (\ref{int1})  can be exploited to
derive an explicit  exact result for $P_S(t)$ for  all $t$. For $d>2$,
we  use a different  method. We  will present  an explicit  result for
$\mu(t)$ in the physically relevant dimension $d=3$, though our method
can, in principle, be used for  any $d>2$.  Note that this problem has
an  alternative  physical  description.   In  terms  of  the  relative
coordinates  ${\vec  R_i}={\vec r_i}-  {\vec  R_0}$,  it represents  a
system of noninteracting particles diffusing in the presence of a constant
drift velocity in  the negative  $z$ direction. The physical origin of 
this drift could be, for example, an external field such as gravity or 
an electric  field. The  survival  probability $P_S(t)$  of  the  tracer
particle, in  this alternative formulation, is  simply the probability
that none of the particles hits the origin up to time $t$.

We consider first the case  $d<2$. The substitution ${\vec R_0}(t)= ct
{\bf {\hat z}}$ in Eq.\ (\ref{int1}) reduces it to a convolution form,
\begin{equation}
\rho    =   \int_0^t    dt'   {\dot    \mu}(t'){    {   \exp\left[-c^2
(t-t')/{4D}\right]} \over { {\left[4\pi D (t-t')\right]}^{d/2} } },
\label{int2}
\end{equation}
which can, subsequently, be solved by the Laplace transform method. We
denote  $\alpha=c^2/4D$   and  define  the   Laplace  transform
${\tilde  \mu}(s)=\int_0^{\infty}  \mu(t)  e^{-st}dt$. Taking  Laplace
transforms on both sides of  Eq.\ (\ref{int2}) and using $\mu(0)=0$, we
obtain ${\tilde \mu}(s)=  A { {(\alpha+s)^{1-d/2}}\over {s^2}}$, where
$A=  \rho(4\pi D)^{d/2}/\Gamma(1-d/2)$  is a  constant. To  invert the
Laplace   transform,  we write   ${\tilde  \mu}(s)=A(1/s+\alpha/
{s^2})(\alpha+s)^{-d/2}$. The inverse Laplace transform of the first
factor, $(1/s+\alpha/s^2)$, is simply $(1+\alpha t)$ and that of the
second  factor,  $(\alpha+s)^{-d/2}$, can  be  easily  found  to  be
$e^{-\alpha t}t^{d/2-1}/{\Gamma(d/2)}$. We  then use the convolution
theorem again to write $\mu(t)$ as
\begin{equation}
\mu(t)= {A\over{\Gamma(d/2)}}\int_0^t \left[1+\alpha(t-t_1)\right]
e^{-\alpha t_1} {t_1}^{d/2-1}dt_1.
\label{mut1}
\end{equation}
The integral on the right-hand side can be expressed in closed form as
\begin{equation}
\mu(t)=B\left[(1+\alpha t)\gamma(d/2,\alpha t)
-\gamma(d/2+1,\alpha t)\right],
\label{mut2}
\end{equation}
where $B=  \rho(4\pi D)^{d/2}\sin (\pi  d/2)/{[\pi \alpha^{d/2}]}$ and
$\gamma(\nu,x)=\int_0^{x}\exp(-y)y^{\nu-1}dy$ is  the incomplete Gamma
function.  Note  that  the   exact  expression for  $\mu(t)$  in  Eq.\
(\ref{mut2}) is valid for all $t$ and $d<2$.

Let  us consider some  limiting cases  of the  general result  in Eq.\
(\ref{mut2}). In the  limit when the velocity $c\to  0$ at  fixed $t$,
our problem  reduces to the  static `target annihilation'  problem for
which  exact results  are  already available\cite{TA,BB1}.   Expanding
Eq.\ (\ref{mut2})  for small $\alpha= c^2/4D$ we  find, to leading
order, $\mu(t)=A_1 t^{d/2}$ where $A_1=2\rho (4\pi D)^{d/2} {\sin (\pi
d/2)}/{\pi d}$.  This indicates a stretched exponential  decay for the
survival  probability  $P_S(t)=\exp(-A_1  t^{d/2})$ and  the  constant
$A_1$  matches   exactly  that   of  the  static   case  derived
earlier\cite{TA,BB1}.  Note that  the limit $c\to 0$ for  fixed $t$ is
equivalent  to  the limit  $t\to  0$ with  $c$  fixed,  since in  Eq.\
(\ref{mut2})  the time  $t$ appears  only in  the  scaling combination
$\alpha t$.   Thus the  result $P_S(t)=\exp(-A_1t^{d/2})$  also holds
for small $t$ with $c$ fixed.  We now consider the opposite limit of 
$t \to  \infty$ at fixed $c$. In this case we find, from
Eq.\ (\ref{mut2}), $\mu(t)\to A_2  t$ as $t\to \infty$ where $A_2=\rho
(4\pi D)^{d/2}  {\sin (\pi d/2)}\Gamma(d/2)/[\pi  \alpha^{(d-2)/2}]$. 
This indicates an exponential decay for the survival probability at late
times,   $P_S(t)\to  \exp(-\theta   t)$  where   the   decay  exponent
$\theta=A_2$ is given by the following exact expression
\begin{equation}
\theta = \rho \pi^{d/2-1} (4D)^{d-1} \sin(\pi d/2)\Gamma(d/2)c^{2-d},
\label{theta1}
\end{equation}
as a function of the $4$ physical parameters $\rho$, $c$, $D$ and $d$.

The marginal dimension $d=2$ is a special case. The integral equation 
(\ref{int1}) is still  valid provided one introduces  an  ultraviolet  
cut-off  reflecting the  necessity  of  a lattice  structure. 
Alternatively, a short time cut-off $t_0$ can be introduced in the 
diffusion propagator: $G\left({\vec R_0}(t), t|{\vec R_0}(t'),t'\right) 
=\exp\{-[{\vec R_0}(t)-{\vec R}(t')]^2/{4D(t-t'+t_0)}\}/[4\pi D (t-t'+t_0)]$, 
where we  consider $t_0$ to  be small. To  extract the leading asymptotic  
behavior, one can  put $t_0=0$  inside the  exponential and  retain it  
only  in the denominator  of  the  propagator. Using  this  propagator  
in  Eq.\ (\ref{int1}), substituting ${\vec R_0}(t)= ct {\bf {\hat z}}$ 
and then taking the Laplace transform as for $d<2$ gives ${\tilde \mu}(s)=
{4\pi\rho D}/[s^2 {\tilde g}(s)]$ where    ${\tilde g}(s)=\int_0^{\infty}  
dt \exp[-(\alpha + s)t]/(t+t_0)$.  Unlike  the
$d<2$   case,   it   is   now   difficult  to   invert   the   Laplace
transform. However, the  large-$t$ behavior of $\mu(t)$  can be easily
extracted  from the  $s\to 0$  behavior of  the Laplace  transform. As
$s\to   0$,   ${\tilde   g}(s)\to   F(\alpha  t_0)$   where   $F(x)=
\int_0^{\infty} du \exp(-u)/(u+x)$. Inverting ${\tilde \mu}(s)$, one
then  gets  $\mu(t)\to  [4\pi   \rho  D/F(\alpha  t_0)]t$  as  $t\to
\infty$.    In    the    limit    $t_0 \ll 1/\alpha$,    one    gets
$F(\alpha t_0)\approx  -\log  (\alpha  t_0)$.  Thus,  the  survival
probability    again    decays    exponentially   for    large    $t$,
$P_S(t)=\exp(-\theta t)$, where $\theta$ is now nonuniversal, $\theta=
4\pi\rho D/[-\log (\alpha t_0)]$.

We turn now to  the  case $d>2$.   We  consider  a spherical  tracer
particle of radius $a$ moving ballistically with constant velocity $c$
in the $z$ direction. Unfortunately, for $d>2$ we do not yet have an
analog of Eq.\ (\ref{int1}). Thus  one has to resort to the original
single particle formulation in Eq.\ (\ref{pst2}). Fortunately, for the
ballistic case in $d>2$, the exact asymptotic behavior of $P_S(t)$ for
large $t$ can  be derived even within this  formulation.  It turns out
to  be   advantageous  in  this  case  to   consider  the  alternative
description  of  the problem  in  terms  of  the relative  coordinates
$R_i(t)=r_i(t)-{\vec R_0}(t)$ where the traps diffuse independently in
the presence of an external drift along the negative $z$ direction. 
Upon shifting to this  relative  coordinate,  one   finds  from   Eq.\
(\ref{pst2}),
\begin{equation}
P_S(t)= \exp\left[-\rho\int P(\vec R,t)d\vec R \right]=\exp[-\mu(t)],
\label{pst3}
\end{equation}
where  $P(\vec R,t)=1-Q(\vec R,t)$  and  $Q(\vec R,t)$  is  the
probability that  a trap, diffusing in the presence of an external drift
and starting  at the initial position $\vec R$ outside  the sphere of
radius  $a$,  does not  hit  the surface  of  the  sphere before  time
$t$. The integral in Eq.\ (\ref{pst3}) is now restricted to the region 
$|\vec R|\geq a$.

It  is easy  to see  that the  probability $Q(\vec  R,t)$  satisfies a
backward Fokker-Planck equation,
\begin{equation}
{{\partial  Q}\over {\partial  t} }=  D {\nabla}^2  Q -  c  {\bf {\hat
z}}{\cdot} {\vec \nabla}Q,
\label{bfp1}
\end{equation}
where the first term on the right-hand  side represents the diffusion
and the second term represents  the drift. This is a backward equation
since we  are varying the initial  position $\vec R$ of  the trap. 
Eq.\ (\ref{bfp1})  holds in the  region $|\vec R|\geq a$  with the
boundary  conditions $Q(\vec  R,t)=0$  for $|\vec  R|=a$ and  $Q(\vec
R,t)\to 1$  as $|\vec R|\to \infty$.  This is because  if the particle
starts at the surface of  the sphere at $t=0$, the probability that it  
does not hit the surface before time $t$ vanishes for all $t>0$. 
Similarly, if the particle starts at infinity, with probability
$1$ it  will not hit the sphere in any finite  time $t$. Evidently
the probability $P(\vec R, t)=1-Q(\vec  R, t)$ also satisfies the same
backward Fokker-Planck equation in  (\ref{bfp1}) (with $Q$ replaced by
$P$) but with a reversal  of boundary conditions, $P(\vec R, t)=1$ for
$|\vec R|=a$ and $P(\vec R, t)\to 0$ as $|\vec R|\to \infty$.

Note that this  Fokker-Planck equation can be written  as a continuity
equation,  ${\partial_t P}  +{\vec \nabla}{\cdot}  {\vec J}=0$  with a
current  $\vec J  = -D  {\vec \nabla}P  + c  {\bf {\hat  z}}P$. Before
solving this  equation, we first  make a simple observation.  We have,
from Eq.\  (\ref{pst3}), $\mu(t)=\rho \int_{|\vec R|\geq  a} P(\vec R,
t)d  {\vec  R}$.  Therefore,  ${\dot \mu}(t)=  \int_{|\vec  R|\geq  a}
{\partial_t  P}\,d {\vec  R}$.  Using the  continuity  equation and 
Gauss's divergence theorem, we obtain
\begin{equation}
{\dot \mu}(t)= -\rho D  \int_{|\vec R|=a} {\vec \nabla}P{\cdot} d{\vec
S},
\label{surface1}
\end{equation}
where  the surface  integral  is over  the  sphere of  radius $a$.  In
deriving Eq.\ (\ref{surface1}) we have used the boundary conditions on
$P$ and  the identity $\int_{|\vec R|=a} {\bf  {\hat z}}{\cdot} d{\vec
S}=0$.

It turns out that, for $d>2$, Eq.\ (\ref{bfp1})  has a stationary
solution. This is because the particle always has a finite probability
to escape the  sphere for $d>2$. Thus, in order to extract the leading
asymptotic  behavior of  $\mu(t)$ for  $t\to \infty$, one  can replace
$P(\vec  R, t)$  by its  stationary solution  $P_{st}(\vec R)$  on the
right  hand side  of Eq.\  (\ref{surface1}). A  subsequent integration
over $t$, using $\mu(0)=0$, shows that, to leading order for large $t$,
$\mu(t) = \theta t$ where
\begin{equation}
\theta= -\rho  D \int_{|\vec R|=a} {\vec  \nabla} P_{st}{\cdot} d{\vec
S}.
\label{surface2}
\end{equation} 
Note that while Eq.\ (\ref{surface1}) is valid for all $d$, the result
in Eq.\ (\ref{surface2}) is valid  only for $d>2$. This is because the
trick  of   replacing  $P(\vec   R,t)$  by  its   stationary  solution
$P_{st}(\vec  R)$  does not  work  for $d\leq  2$  since  there is  no
stationary solution in that case.

Thus for $d>2$ the survival probability also decays exponentially for
large  $t$, $P_S(t)\sim  \exp(-\theta t)$  with the  exponent $\theta$
given by  the general formula in Eq.\  (\ref{surface2}).  Obtaining an
explicit  expression  for  $\theta$  requires  knowledge  of  the
stationary solution of Eq.\ (\ref{bfp1})  which we now provide for the
physically relevant dimension $d=3$. The stationary solution satisfies
the equation
\begin{equation}
D{\nabla^2 P_{st}} = c {\bf {\hat z}}{\cdot}{\vec \nabla}P_{st},
\label{stat1}
\end{equation}
with the boundary conditions $P_{st}(\vec R)=1$ for $|\vec R |=a$ and
$P_{st}(\vec R)\to 0$  as $|\vec R|\to \infty$. This  problem does not
have a radial symmetry.  Fortunately, the substitution $P_{st}(\vec R)
=  \exp(\beta z)\psi(\vec  R)$, with $\beta=c/2D$, restores  the radial  
symmetry  since $\psi(\vec R)$ satisfies the Poisson equation $\nabla^2 
\psi = \beta^2 \psi$. The general solution satisfying the boundary 
condition at infinity can be obtained by standard techniques, to give
\begin{equation}
P_{st}(\vec  R)= R^{-1/2}{e^{\beta  R  \cos \phi}}\sum_{l=0}^{\infty}
b_l P_l(\cos \phi) K_{l+{1\over {2}}}(\beta R),
\label{sol1}
\end{equation}
where $R=|\vec R|$,  $\phi$ is the angle between  $\vec R$ and
the $z$  axis, $P_l(x)$ is the  Legendre polynomial of  degree $l$ and
$K_{\nu}(x)$  is the  modified  Bessel function  of  index $\nu$.  The
unknown  coefficients   $b_l$  are  determined   from  the  boundary
condition $P_{st}(R=a)=1$. Substituting $R=a$ in Eq.\ (\ref{sol1}) and
using the  orthogonality properties of the functions $P_l(x)$ gives, 
after some algebra,
\begin{equation}
P_{st}(\vec    R)=    \sqrt{    {a\over    {R}}}e^{\beta    R    \cos
\phi}\sum_{l=0}^{\infty}   (l+{1\over   {2}})   a_l  P_l(\cos   \phi){
{K_{l+{1\over {2}}}(\beta R)}\over {K_{l+{1\over {2}}}(\beta a)}},
\label{sol2}
\end{equation}
where   $a_l=   \int_{-1}^{1}P_l(x)   e^{-\beta   a   x}dx   =   (-1)^l
\sqrt{2\pi/{\beta a}}\, I_{l+1/2}(\beta a)$.

Substituting the  stationary solution, Eq.\ (\ref{sol2}), into Eq.\
(\ref{surface2})  and  performing  the  surface integral,  we  finally
obtain, after a few steps  of algebra, the following rather nontrivial
expression for  $\theta$ in terms  of the physical  parameters $\rho$,
$D$, $a$ and $c$:
\begin{equation}
\theta=   2\pi  a   \rho   D\left[1-2\pi  \sum_{l=0}^{\infty}   (-1)^l
(l+{1\over  {2}}){{K'_{l+{1\over  {2}}}(\beta  a)}\over  {K_{l+{1\over
{2}}}(\beta a)}} I_{l+{1\over {2}}}^2 (\beta a)\right],
\label{theta3}
\end{equation}
where $K'_{\nu}(x)=dK_{\nu}(x)/dx$ and $\beta=c/{2D}$. The series in 
Eq.\ (\ref{theta3}) can be summed numerically. If $H(\beta a)$ is the  
function in the square brackets, then $H(x)$ is monotically 
increasing, with $H(0)=2$ and $H(x)/x \to 1$ for $x \to \infty$. 
The former result, corresponding to $c=0$, recovers the known result,  
$\theta = 4\pi a\rho D$, for a static target \cite{BB2}. The latter, 
corresponding to $c \to \infty$, can be understood by noting that in 
this limit the probability that the sphere has not been hit by a trap 
is given by $\exp(-\rho V)$, where $V=\pi a^2 ct$ is the volume swept 
out by the sphere in time $t$ and we require that this volume initially 
contains no traps.  Hence $\theta \to \pi a^2 c\rho$ in this limit, 
corresponding to $H(x) \to x$. 

In summary,  we have  studied the general  problem of  calculating the
survival probability of a tracer particle moving along a deterministic
trajectory in the presence of diffusing traps. In particular, when the
tracer particle  moves ballistically  we have shown  that its survival
probability  $P_S(t)\sim \exp(-\theta t)$ for  large  $t$  in  all
dimensions.  We  have  derived  exact  expressions  for  the  exponent
$\theta$  in terms  of the  system parameters  for $d\leq  2$  and for
$d=3$.

\end{multicols}


\begin{references}     

\bibitem{Rice} S.A.\ Rice, {\em Diffusion-Limited Reactions} (Elsevier,
Amsterdam, 1985).

\bibitem{Benson}  S.W.\  Benson,   {\em  The  Foundations  of  Chemical
Kinetics} (McGraw-Hill, New York, 1960).

\bibitem{OZ} A.A.\  Ovchinikov and Ya.\ B.\ Zeldovitch,  Chem.\ Phys.\ 
{\bf 28}, 214 (1978).

\bibitem{TW} D.\ Toussaint and  F.\ Wilczek, J.\ Chem.\ Phys.\ {\bf 78},
2642 (1983).

\bibitem{PP} S.\ Redner and P.L.\ Krapivsky, Am.\ J.\ Phys.\ {\bf 67}, 
1277 (1999); S.\  Redner, {\em  A  Guide to  First-passage Processes}  
(CUP, Cambridge, 2001).

\bibitem{WP}  C.\ Monthus, Phys.\ Rev.\ E {\bf  54},  4844  (1996);
S.N.\ Majumdar  and  S.J.\ Cornell, Phys.\ Rev.\  E  {\bf  57},  3757
(1998); S.J.\ O'Donoghue  and A.J.\ Bray,  Phys.\ Rev.\ E {\bf  65}, 
051114 (2002).

\bibitem{TA} M.\ Tachiya, Radiat.\ Phys.\ Chem.\ {\bf 21}, 167 (1983);
A.\ Blumen, G.\ Zumofen, and J.\ Klafter, Phys.\ Rev.\ B {\bf  30}, 5379
(1984); S.F.\ Burlatsky and A.A.\ Ovchinikov, Sov.\ Phys.\ JETP {\bf 65},
908 (1987).

\bibitem{BL} M.\ Bramson and J.L.\ Lebowitz, Phys.\ Rev.\ Lett.\ 
{\bf 61}, 2397 (1988).

\bibitem{MG} V. Mehra  and P. Grassberger,  Phys.\ Rev.\  E {\bf  65},
050101 (2002).

\bibitem{BB1} A.J.\  Bray and R.A.\  Blythe, Phys.\ Rev.\ Lett.\ {\bf 89},
150601 (2002);  R.A.\ Blythe  and A.J.\ Bray, J.\ Phys.\ A  Math-Gen {\bf
35}, 10503 (2002).

\bibitem{BB2} R.A.\ Blythe and A.J.\ Bray, Phys.\ Rev.\ E {\bf 67}, 
041101 (2003).  

\bibitem{OBCM} G.\ Oshanin, O.\ B\'enichou,  M.\ Coppey, and  M.\ Moreau,
Phys.\ Rev.\ E {\bf 66}, 060101 (2002); see also cond-mat/0303273.

\bibitem{BMB} A.J.\ Bray, S.N.\ Majumdar, and R.A.\ Blythe, cond-mat/0212226 
(to appear in Phys. Rev. E).

\bibitem{Sal} P.\ Salminen, Adv.\ Appl.\ Prob.\ {\bf 20}, 411 (1988).

\end{references}
\end{document}